# GeP$_3$: A small indirect band gap 2D crystal with high carrier mobility and strong interlayer quantum confinement

Yu Jing[1,3], Yandong Ma[1], Yafei Li[2,*], Thomas Heine[1,3]*

1. Wilhelm-Ostwald-Institute for Physical und Theoretical Chemistry, Leipzig University, Linnéstr. 2, 04103 Leipzig, Germany;

2. College of Chemistry and Materials Science, Jiangsu Key Laboratory of Biofunctional Materials, Nanjing Normal University, Nanjing, Jiangsu 210023, China;

3. Department of Physics and Earth Sciences, Jacobs University Bremen, Campus Ring 1, 28759 Bremen, Germany.

*Correspondence and requests for materials should be addressed to Y. L. (email: liyafei @njnu.edu.cn) or to T. H. (email: thomas.heine@uni-leipzig.de)

**Table Of Contents graphic**

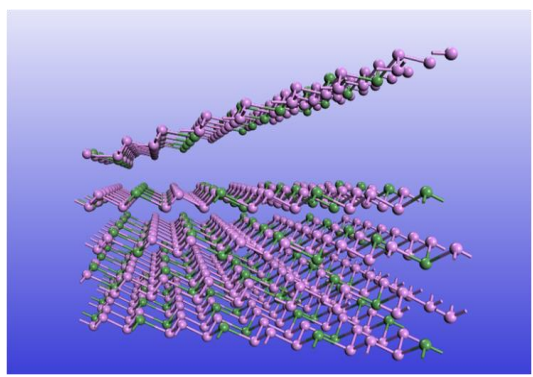

Cleavage of GeP$_3$ monolayer from the bulk

**Abstract**

We propose a two-dimensional crystal which possesses low indirect band gaps of 0.55 eV (monolayer) and 0.43 eV (bilayer) and high carrier mobilities similar to those of phosphorene: $GeP_3$. $GeP_3$ has a stable three-dimensional layered bulk counterpart which is metallic and is known from experiment since 1970. $GeP_3$ monolayer has a calculated cleavage energy of 1.14 J m$^{-2}$, which suggests exfoliation of bulk material as viable means for the preparation of mono- and few-layer materials. The material shows strong interlayer quantum confinement effects, resulting in a band gap reduction from mono- to bilayer, and then to a semiconductor-metal transition between bi- and triple layer. Under biaxial strain, the indirect band gap can be turned into a direct one. Pronounced light absorption in the spectral range from ~600 to 1400 nm is predicted for monolayer and bilayer and promises applications in photovoltaics.

**Introduction**

Since the discovery of graphene,[1-3] the family of two-dimensional (2D) crystals[4] has grown considerably, encompassing today a rich variety offering almost all desirable electronic properties[5] that are required for nanoelectronics. 2D Dirac semimetals (e.g. graphene, silicene, germanene and stanene),[6-11] insulators (e.g. h-BN),[12-14] direct band-gap semiconductors (e.g. transition metal dichalcogenides (TMDC) and phosphorene)[15-20] are readily available. First applications in optoelectronics have been reported on the basis of TMDC,[21,22] however, the relatively large band gap of nearly 2 eV does not allow efficient collection of the solar emission. The most commonly used solar cells are based on silicon and group III-V semiconductors (e.g. GaAs, InAs, InP), all of which have band gaps below 1.50 eV. To this end, 2D semiconductors with a band gap in the range of 0.3~1.5 eV are desirable, however, to date they are rarely known from experiment.

Monolayers of Group 14 elements (C, Si, Ge, Sn) are semimetallic, while phosphorene has a wide band gap. Combining a Group 14 element with P can be expected to result in a material with moderate band gap. Indeed, recently Guan *et al.* proposed theoretically a new 2D structure, phosphorus carbide (PC).[23] The PC monolayer is reported to be semiconducting with a relatively narrow band gap of ~0.7 eV. However, as we know no corresponding layered bulk phase of PC from nature, the experimental realization of PC monolayer will probably be rather challenging.

On the other hand, a layered material composed of P and Ge with stoichiometry $GeP_3$ has already been reported in the 1970´s and its synthesis is well-known.[24,25] Generally, three phases of $GeP_x$ ($x$= 1, 3 or 5), where $x$ is controlled by the reaction conditions, have been realized in experiment,[24,25] and clean-phase single crystals can be obtained. According to these early experimental reports, $GeP_3$ crystal possesses the puckered arsenic-type honeycomb structure in ABC stacking, is superconductive and, importantly, crystallizes in the layered structure[24,25] shown in Figure 1a. According to Hulliger, the replacement of every fourth P atom by Ge introduces an additional valency, which enforces interlayer interaction and is responsible for the metallic

character of bulk GeP$_3$.[26] Exfoliation to the monolayer should, however, be a reachable goal.[27-29]

Here, by means of first principles calculations, we study stability, mechanical, electronic and optoelectronic properties of GeP$_3$ in detail, ranging from the monolayer (1L) via bi- (2L) triple- (3L) and quadruple (4L) layers to the layered bulk. We show that 1L and 2L GeP$_3$ are chemically, mechanically and dynamically stable. GeP$_3$ shows remarkably strong interlayer interactions, which result in different electronic characteristics for different layer number. 1L GeP$_3$ possesses a moderate indirect band gap of 0.55 eV, 2L GeP$_3$ has a smaller band gap of 0.43 eV and shows remarkably high carrier mobilities of ~8.84 ×10$^3$ cm$^2$ V$^{-1}$s$^{-1}$ and ~8.48 ×10$^3$ cm$^2$ V$^{-1}$s$^{-1}$ for electrons and holes, respectively. Moreover, it shows strong light absorption in the visible and infrared regions, making it a particularly strong candidate as component in ultrathin and flexible solar cells. 3L GeP$_3$ and any thicker layer formation are metallic as the bulk.

**Results and Discussion**

*GeP$_3$ layered 3D crystal*

Based on our DFT calculations, the lattice parameters of bulk GeP$_3$ with space group $R\bar{3}m$ were optimized to be a = b = 7.09 Å and c = 9.62 Å, which are in good accordance with the results of the experimental study (a = b = 7.05 Å, c = 9.93 Å).[25] As shown in Figure 1 a and b, each Ge atom forms three Ge-P bonds with three neighboring P atoms, and each P atom forms two P-P bonds and one Ge-P bond with neighboring P and Ge atoms, respectively.

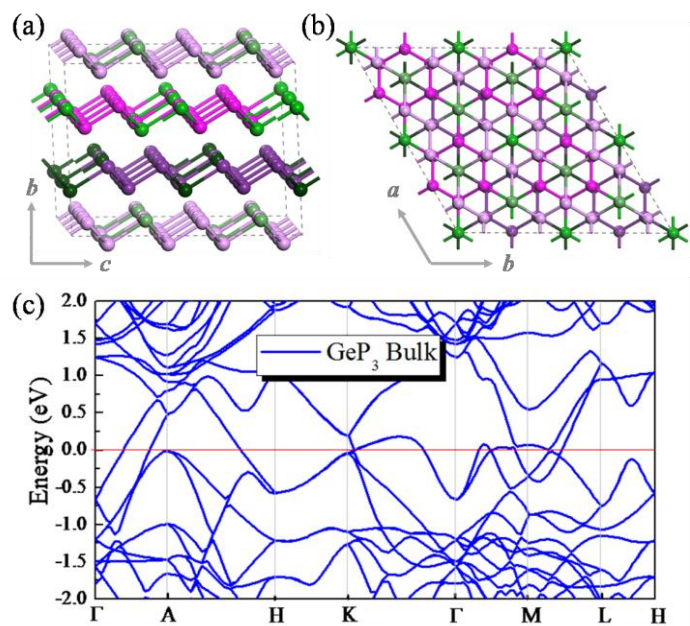

Figure 1. Structure of bulk GeP$_3$ in a 2×2×1 supercell from side (a) and top (b) views. Pink and green balls represent P and Ge atoms, respectively. (c) Calculated band structure of bulk GeP$_3$.

In each layer of GeP$_3$, the atoms are located in two hexagonal planes in close proximity, forming a puckered single-layered structure (Figure 1a) akin to that of germanene and blue phosphorene.[30-32] The lattice parameters, bond length and bond angles for bulk GeP$_3$ obtained from DFT calculations are shown in Table S1 and compared with those obtained from experimental studies. GeP$_3$ bulk crystal is metallic with bands spreading across the Fermi level, as shown in Figure 1c.

*Structure and Stability of 2D GeP$_3$*

1L GeP$_3$ has been created by full optimization of a single layer taken out from the bulk structure. As shown in Figure 2a, its structural characteristics remain and it exhibits a hexagonal honeycomb configuration. However, the puckering of 1L GeP$_3$ (Figure 2b) is more pronounced than that of the bulk, which is reflected in a lattice shrinkage of 1.8%, and results in optimized lattice parameters of a = b = 6.96 Å. This lattice shrinkage contributes to the bond length and bond angle change from bulk to monolayer, as illustrated in Table S1. It is also the driving force for the semiconducting properties of 1L GeP$_3$, as the unoptimized monolayer as taken from the bulk is metallic.

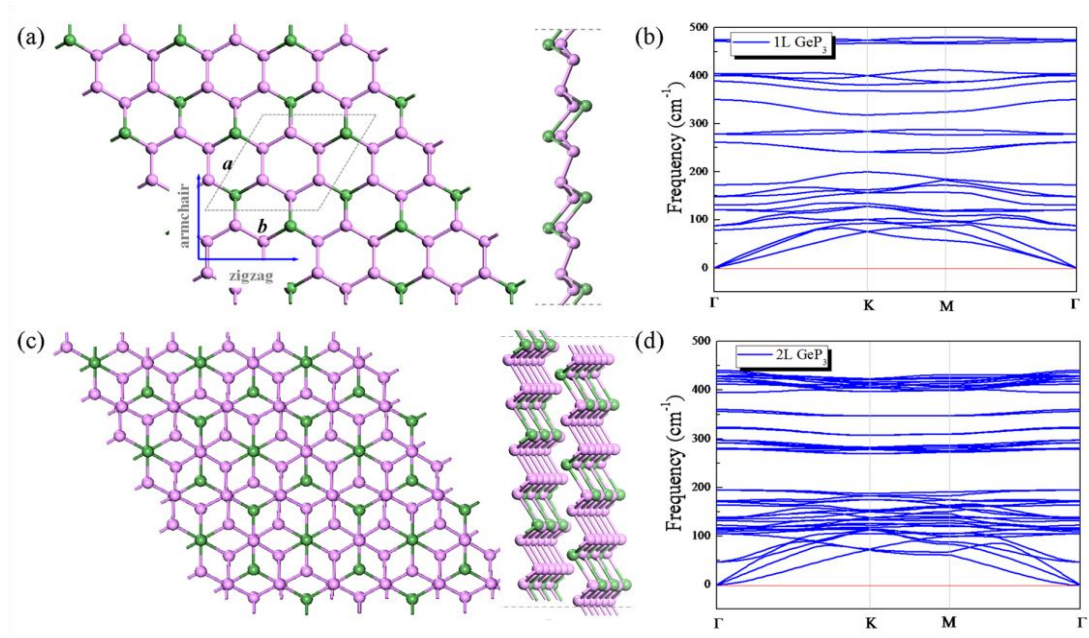

Figure 2. Structure of (a) 1L GeP$_3$ (top and side views), the hexagonal unit cell is enclosed in the grey dashed line, and (b) its phonon spectrum; (c) structure and (d) phonon spectrum of 2L GeP$_3$.

The phonon dispersion curves of 1L GeP$_3$, shown in Figure 2b, contain only real modes in the form typical for 2D crystals (one parabolic and two linear acoustic branches starting from the Γ point), which suggests kinetic stability. The highest frequency mode of 1L GeP$_3$ reaches 480 cm$^{-1}$, which is very close to that of MoS$_2$ (473 cm$^{-1}$) and comparable to silicene (580 cm$^{-1}$),[33,34] indicating the mechanical robustness of the covalent P-P bonds. If compared to phosphorene, we note that the highest frequency mode of GeP$_3$ is higher than that of phosphorene (~450 cm$^{-1}$).[35]

The kinetic and thermal stability is further substantiated by *ab* initio molecular dynamics (AIMD) simulations (Figure S2), where the 1L GeP$_3$ structure remains intact at 500K after a 10 ps, and subsequent geometry optimization recovers the initial structure. 2L GeP$_3$ in AB stacking order (as found in bulk, Figure 2c) has a similar lattice constant (a=b=6.95 Å) as the monolayer, however, P-P bonds slightly enlarge, while Ge-P bonds shrink (Table S1). The P-P bond length increase is reflected by a lowering of the highest phonon frequencies (Figure 2d). As no imaginary phonon mode is observed, we conclude that free-standing 2L and 1L GeP$_3$ are stable 2D crystals.

Moreover, we expect higher chemical stability for 1L GeP$_3$ compared to phosphorene, as its work function of 4.89 eV is substantially higher (4.25 eV for phosphorene, both values at the PBE level).

*Cleavage energies of GeP$_3$ crystal*

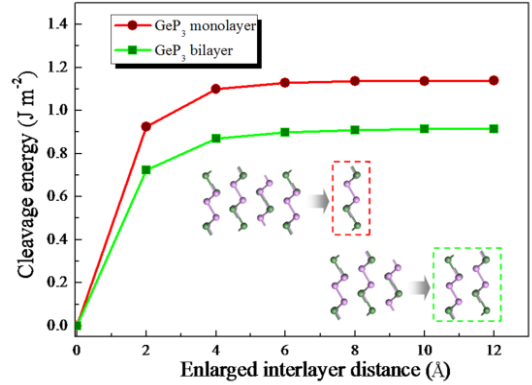

Figure 3. Cleavage energy estimation for the formation of 1L and 2L GeP$_3$, calculated by enlarging the interlayer distance between the 1L/2L system that is removed from the remainder of a 5L slab, resembling the bulk model.

The most popular techniques to produce single and few layer flakes from the layered bulk materials are mechanical cleavage 4 and liquid phase exfoliation.[27] While the latter technique is suitable even for layered materials with strong interlayer interaction energy,[36] the former one requires weak interlayer binding. To assess the possibility of mechanical or liquid phase exfoliation we estimated the cleavage energy of 1L and 2L GeP$_3$ from a 5L slab, serving as a model of the bulk. As shown in Figure 3, increasing the distance between the exfoliating 1L (and 2L) and the bulk, the energy increases, reaching convergence at about 4 Å. The cleavage energy for 1L and 2L GeP$_3$ are ~1.14 J m$^{-2}$ and 0.91 J m$^{-2}$, respectively. For comparison, the experimentally estimated exfoliation energy of graphene is 0.37 J m$^{-2}$,[37] and the DFT estimated exfoliation energy for some layered materials, like Ga$_2$N,[38] NaSnP,[39] and GeS$_2$[40] are 1.09 J m$^{-2}$, 0.81 J m$^{-2}$ and 0.52 J m$^{-2}$, respectively, and thus in the same range as 1L and 2L GeP$_3$.

*Electronic properties of GeP$_3$ thin layers*

The band structure of 1L GeP$_3$ was computed at the PBE level of theory, which is known to produce correct band shapes, but underestimates the band gap. As shown in Figure 4a, at this level of theory 1L GeP$_3$ is semiconducting with an indirect band gap of 0.27 eV. The conduction band minimum (CBM) locates at the Γ point, while the valence band maximum (VBM) locates at the K point. Substantiating the results using the more elaborate Herd−Scuseria−Emzerhof hybrid functional (HSE06) yields similar band structures, but with wider band gap of 0.55 eV (Figure 4a). Due to the (expectedly) strong band gap opening given by the more elaborate HSE06 functional we will adopt this level of theory in all forthcoming band structure calculations.

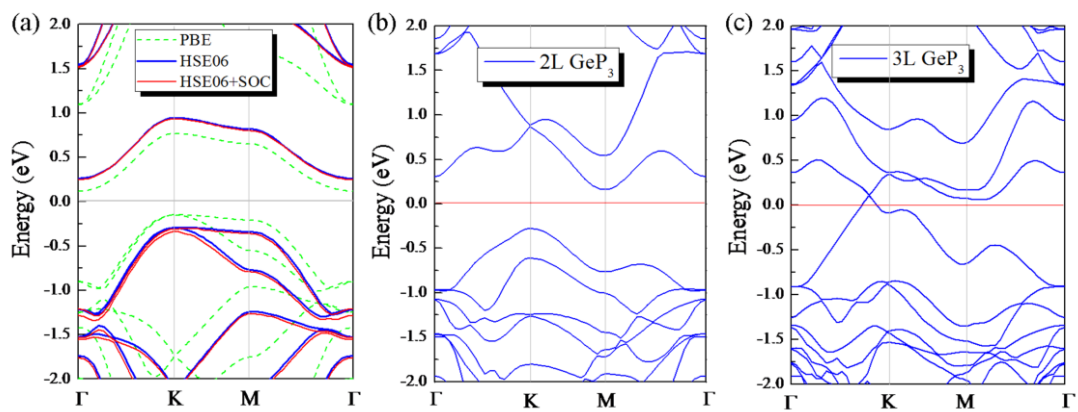

Figure 4. Band structures of 1L GeP$_3$ (a) calculated at PBE, HSE06 and HSE06+SOC levels of theory, and 2L (b) and 3L (c) GeP$_3$, calculated at the HSE06 level.

Since Ge is a relatively heavy element, spin-orbital-coupling (SOC) may influence its electronic properties, even though no major effects are expected due to the inversion symmetry of the crystals. As shown in Figure 4a, there is only a slight energy level splitting around the Γ point, and overall SOC has a negligible influence on the band structure, in particular concerning band gap and frontier bands. This can be ascribed to the fact that the amount of Ge in GeP$_3$ is only 25 at.% and the wavefunctions and partial density of states (PDOS), as shown in Figures S2 and S3a, indicate that the main contributions to the VBM and CBM can be attributed to the P-$2p$ orbitals. Thus, the SOC effect is not pronounced enough to significantly influence the band structure, and hereafter we will neglect SOC unless stated otherwise.

The interesting fact that 1L GeP$_3$ becomes semiconducting after isolation from the metallic bulk indicates particularly strong quantum confinement effects in GeP$_3$.

Therefore, we studied the electronic properties of $n$L GeP$_3$ systems, with $n$=2, 3, 4, which adopt the same stacking order as in the bulk. The band structure of 2L GeP$_3$ (Figure 4b) characterises the material being a semiconductor with indirect band gap of 0.43 eV. In contrast to 1L GeP$_3$, VBM and CBM of 2L GeP$_3$ locate at the K and M points, resepctively.

The band structure of 3L GeP$_3$ (Figure 4c) shows a semiconductor-metal transition, and also 4L GeP$_3$ remains metallic (Figure S3b). Thus, $n$L GeP$_3$ show a remarkable diversity in electronic structure, ranging from low-band gap semiconductors to metals (Figure S4). If the layer thickness can be controlled in experiment, all-in-one devices could be produced, e.g. field effect transistors where electrodes and semiconductor are made out of the same material, as already suggested for MoS$_2$[41] and PdS$_2$.[42]

To better understand the electronic properties of 2D GeP$_3$ and their potential for electronic applications, we calculated the carrier mobilities of 1L and 2L GeP$_3$. If the phonon scattering dominates the intrinsic mobility, as it is typically the case for inorganic semiconductors, the carrier mobility can be calculated by utilizing deformation potential (DP) theory,[43] which has turned out to be effective in predicting the carrier mobilities of many 2D semiconductors, including, but not limited to, phosphorene and 1L MoS$_2$.[44,45] The details for the carrier mobility calculations are provided in the SI (Figure S5, S6 and Table S2).

Our calculations indicate that 2L GeP$_3$ possesses higher carrier mobilities compared to its 1L counterpart, which was expected already from the band structure (smaller effective mass in 2L) and phonon dispersion relation (higher stiffness for 2L). The carrier mobilities of 2L GeP$_3$ can be as high as $8.84 \times 10^3$ cm$^2$ V$^{-1}$s$^{-1}$ for electrons and $8.48 \times 10^3$ cm$^2$ V$^{-1}$s$^{-1}$ for holes if the current flows along the armchair direction. However, mobilities are strongly direction dependent, as the value decreases to $1.25 \times 10^3$ cm$^2$ V$^{-1}$s$^{-1}$ for electrons and $4.63 \times 10^3$ cm$^2$ V$^{-1}$s$^{-1}$ for holes along the zigzag direction, which are, however, still comparable to those of phosphorene (~$10^4$ cm$^2$ V$^{-1}$ s$^{-1}$).[44] Since it has been indicated that due to the lower cleavage energy, 2L GeP$_3$ will probably be produced in higher yields by exfoliation of the bulk, we believe that 2L GeP$_3$ bilayer is a promising material for application in 2D electronics.

We also studied the effects of compressive and tensile biaxial strain on the band structure of 1L GeP$_3$, since applying elastic strain can be an effective means of band structure engineering in 2D semiconductors. As shown in Figure S7a, the band gap of 1L GeP$_3$ increases with increasing tensile strain, and decreases upon compression. When the applied tensile strain is 5%, the band gap of 1L GeP$_3$ (0.55 eV) is enlarged to be 0.73 eV. When the compressive strain is 5%, the band gap is reduced to 0.21 eV. At compressive biaxial strain of 5%, the VBM of GeP$_3$ monolayer moves from the K point to the Γ point (Figure S7b), resulting in an indirect-to-direct band gap transition. Therefore, our investigations indicate that the electronic properties of GeP$_3$ monolayer can be effectively tuned by applying external strain, which could lead to wide applications in flexible electronics.

*Optical properties*

As low band gap semiconductors can absorb light in the visible range, we estimate the light-harvesting performance of 1L and 2L GeP$_3$ by calculating their absorption spectra in- and out-of-plane using the HSE06 functional.

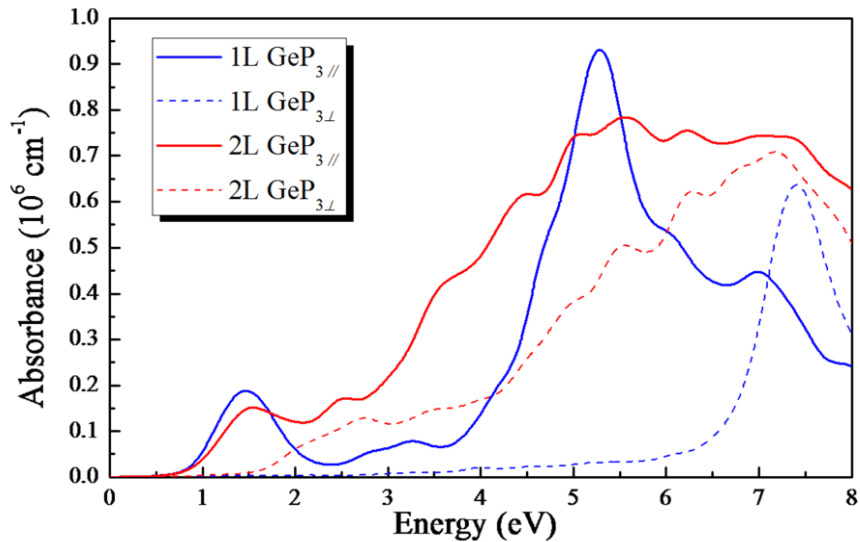

Figure 5. Calculated absorbtion spectra for 1L and 2L GeP$_3$ from the parallel (∥) and perpendicular (⊥) directions of the 2D structures at the HSE06 level.

As shown in Figure 5, the absorption coefficients of 1L and 2L GeP$_3$ reach to the order of 10$^6$ cm$^{-1}$, which are comparable to those of organic perovskite solar cells.[46,47]

In-plane absorption is always larger (due to the larger cross section), suggesting that flakes should be aligned normal to the surface of the photovoltaic cell for most efficient application. The bilayer shows excellent absorption, in particular in the important region between 1 and 4 eV, which marks the infrared, visible and near UV range of the solar spectrum.

**Conclusion**

In summary, we have shown that 2D crystals $n$L GeP$_3$, with layer numbers $n$=1-3, are remarkable new candidates for the realization of interesting nanoelectronic and optoelectronic devices. Synthesis of the metallic layered bulk species is known since 1970, and the predicted cleavage energies indicate that exfoliation from the bulk is possible. The material shows a remarkable interlayer quantum confinement, resulting in semiconducting mono- and bilayer structures, while three and more layers result in metallic species. Mono-and bilayer show small indirect band gaps (0.55 eV and 0.43 eV, respectively), and the latter has electron and hole mobilities comparable to those of phosphorene. An indirect-to-direct band gap transition is possible for the monolayer by strain engineering. In addition to extraordinary high carrier mobilities, GeP$_3$ bilayer has a pronounced light absorption in the range of the solar spectrum, and thus is promising for solar cell applications. Besides, the presence of metallic and semiconducting 2D species made from the same bulk material, only differing in the number of layers, offers interesting new device concepts which employ electrodes and semiconductor made of GeP$_3$.

**Methods**

The hexagonal lattices of GeP$_3$ were optimized using density-functional theory (DFT) within the projected-augmented wave (PAW) method[48,49] and a plane wave basis as implemented in Vienna ab initio simulation package (VASP).[50] The generalized gradient approximation (GGA) as suggested by Perdew, Burke and Ernzerhof with London dispersion corrections as proposed by Grimme (PBE+D2) was employed to accurately describe the weak interactions between GeP$_3$ layers.[51,52] Geometry optimizations were performed with a convergence threshold of 10$^{-4}$ eV in energy and $10^{-2}$ eV/Å for the force by using the conjugated gradient method. A cutoff energy of 500 eV for the plane-wave basis set was adopted in all computations.

**Supporting Information Available:**

Details of the DFT calculation, structural description for $GeP_3$ bulk and thin layers, wavefunctions and PDOS of 1L $GeP_3$, band structure of 4L $GeP_3$ and details for the calculations of the carrier mobilities according to the DP theory and band structure engineering by strain. This material is available free of charge via the Internet at http://pubs.acs.org.

**Acknowledgement**

We thank Dr. Agnieszka Kuc for valuable discussion. Financial support by the Natural Science Foundation of China (No. 21522305 and 21403115), the NSF of Jiangsu Province of China (No. BK20150045), and by FlagERA (DFG HE 3543/27-1) is gratefully acknowledged. We thank ZIH Dresden for supercomputing resources.

The authors declare no competing financial interests.

**References**

1  Novoselov, K. S.; Geim, A. K.; Morozov, S. V.; Jiang, D.; Zhang, Y.; Dubonos, S. V.; Grigorieva, I. V.; Firsov, A. A. *Science* **2004**, *306*, 666 − 669.

2  Allen, M. J.; Tung, V. C.; Kaner, R. B. *Chem. Rev.* **2010**, *110*, 132 − 145.

3  Rao, C. N. R.; Sood, A. K.; Subrahmanyam, K. S.; Govindaraj, A. *Angew. Chem. Int. Ed.* **2009**, *48*, 7752 − 7778.

4  Novoselov, K. S.; Jiang, D.; Schedin, F.; Booth, T. J.; Khotkevich, V. V.; Morozov, S. V.; Geim, A. K. *Proc. Natl. Acad. Sci. U. S. A.* **2005**, *102*, 10451 − 10453.

5  Miro, P.; Audiffred, M.; Heine, T. *Chem. Soc. Rev.* **2014**, *43*, 6537 − 6554.

6  Fleurence, A.; Friedlein, R.; Ozaki, T.; Kawai, H.; Wang, Y.; Yamada-Takamura, Y. *Phys. Rev. Lett.* **2012**, *108*, 245501.

7  Vogt, P.; Padova, P. D.; Quaresima, C.; Avila, J.; Frantzeskakis, E.; Asensio, M. C.; Resta, A.; Ealet, B.; Lay, G. L. *Phys. Rev. Lett.* **2012**, *108*, 155501.

8  Dávila, M. E.; Xian, L.; Cahangirov, S.; Rubio, A.; Lay, G. L. *New J. Phys.* **2014**, *16*,

095002.

9   Li, L.; Lu, S.; Pan, J.; Qin, Z.; Wang, Y.; Wang, Y.; Cao, G.; Du, S.; Gao, H. *Adv. Mater.* **2014**, *26*, 4820 – 4824.

10  Derivaz, M.; Dentel, D.; Stephan, R.; Hanf, M.-C.; Mehdaoui, A.; Sonnet, P.; Pirri, C. *Nano Lett.* **2015**, *15*, 2510 − 2516.

11  Zhu, F.; Chen, W.; Xu, Y.; Gao, C.; Guan, D.; Liu, C.; Qian, D.; Zhang, S.-C.; Jia, J. *Nat. Mater.* **2015**, *14*, 1020 – 1025.

12  Zhi, C. Y.; Bando, Y.; Tang, C. C.; Kuwahara, H.; Golberg, D. *Adv. Mater.* **2009**, *21*, 2889 − 2893.

13  Warner, J. H.; Rümmeli, M. H.; Bachmatiuk, A.; Büchner, B. *ACS Nano* **2010**, *4*, 1299 − 1304.

14  Nag, A.; Raidongia, K.; Hembram, K. P. S. S.; Datta, R.; Waghmare, U. V.; Rao, C. N. R. *ACS Nano* **2010**, *4*, 1539 − 1544.

15  Ramakrishna Matte, H. S. S.; Gomathi, A.; Manna, A. K.; Late, D. J.; Datta, R.; Pati, S. K.; Rao, C. N. R. *Angew. Chem. Int. Ed.* **2010**, *49*, 4059 − 4062.

16  Pizzi, G.; Gibertini, M.; Dib, E.; Marzari, N.; Iannaccone, G.; Fiori, G. *Nat. Comm.* **2016**, *7*, 12585.

17  Mak, K. F.; Lee, C.; Hone, J.; Shan, J.; Heinz, T. F. *Phys. Rev. Lett.* **2010**, *105*, 136805.

18  Wang, Q. H.; Kalantar-Zadeh, K.; Kis, A.; Coleman, J. N.; Strano, M. S. *Nat. Nanotech.* **2012**, *7*, 699 – 712.

19  Li, L.; Yu, Y.; Ye, G. J.; Ge, Q.; Ou, X.; Wu, H.; Feng, D.; Chen, X. H.; Zhang, Y. *Nat. Nanotech.* **2014**, *9*, 372 − 377.

20  Liu, H.; Neal, A. T.; Zhu, Z.; Luo, Z.; Xu, X.; Tomanek, D.; Ye, P. D. *ACS Nano* **2014**, *8*, 4033 − 4041.

21  Lopez-Sanchez, O.; Lembke, D.; Kayci, M.; Radenovic, A.; Kis, A. *Nat. Nanotech.* **2013**, *8*, 497 − 501.

22  Jaegermann, W.; Klein, A.; Mayer, T. *Adv. Mater.* **2009**, *21*, 4196 − 4206.

23  Guan, J.; Liu, D.; Zhu, Z.; Tománek, D. *Nano Lett.* **2016**, *16*, 3247 – 3252.

24  Gullman, J.; Olofsson, O. *J. Solid State Chem.* **1972**, *5*, 441 − 445.

25  Donohue, P. C.; Young, H. S. *J. Solid State Chem.* **1970**, *1*, 143 − 149.

26  Hulliger F. Structural chemistry of layer-type phases [M]. Springer Science & Business Media, 2012.

27  Nicolosi, V.; Chhowalla, M.; Kanatzidis, M. G.; Strano, M. S.; Coleman, J. N. *Science*, **2013**, *340*, 6139.

28  Paton, K. R.; Varrla, E.; Backes, C.; Smith, R. J.; Khan, U.; O'Neill, A.; Higgins, T. *Nat.*

*Mater*. **2014**, *13*, 624 − 630.

29  Hanlon, D.; Backes, C.; Doherty, E.; Cucinotta, C. S.; Berner, N. C.; Boland, C.; Zhang, S. *Nat. Commun.* **2015**, *6*, 8563.

30  Ni, Z.; Liu, Q.; Tang, K.; Zheng, J.; Zhou, J.; Qin, R.; Gao, Z.; Yu, D.; Lu, J. *Nano Lett.* **2012**, *12*, 113 − 118.

31  Roome, N. J.; Carey, J. D. *ACS Appl. Mater. Interfaces* **2014**, *6*, 7743 – 7750.

32  Zhu, Z.; Tománek, D. *Phys. Rev. Lett.* **2014**, *112*, 176802.

33  Molina-Sánchez, A.; Wirtz, L. *Phys. Rev. B* **2011**, *84*, 155413.

34  Cahangirov, S.; Topsakal, M.; Akturk, E.; Sahin, H.; Ciraci, S. *Phys. Rev. Lett.* **2009**, *102*, 236804.

35  Qin, G.; Yan, Q.- B.; Qin, Z.; Yue, S.- Y.; Hu, M.; Su, G. *Phys. Chem. Chem. Phys.* **2015**, *17*, 4854 – 4858.

36  Druffel, D. L.; Kuntz, K. L.; Woomer, A. H.; Alcorn, F. M.; Hu, J.; Donley, C. L.; Warren, S. C. *J. Am. Chem. Soc.* **2016**, *138*, 16089 – 16094.

37  Zacharia, R.; Ulbricht, H.; Hertel, T. *Phys. Rev. B: Condens. Matter Mater. Phys.* **2004**, *69*, 155406.

38  Zhao, S.; Li, Z.; Yang, J. *J. Am. Chem. Soc.* **2014**, *136*, 13313 − 13318.

39  Jiao, Y.; Ma, F.; Gao, G.; Bell, J.; Frauenheim T.; Du, A. *J. Phys. Chem. Lett.* **2015**, *6*, 2682 − 2687.

40  Li, F.; Liu, X.; Wang Y.; Li, Y. *J. Mater. Chem. C* **2016**, *4*, 2155 − 2159.

41  Yamaguchi, H.; Blancon, J.-C.; Kappera, R.; Lei, S.; Najmaei, S.; Mangum, B. D.; Gupta, G.; Ajayan, P. M.; Lou, J.; Chhowalla, M.; Crochet, J. J.; Mohite, A. D. *ACS Nano* **2015**, *9*, 840 − 849.

42  Ghorbani-Asl, M.; Kuc, A.; Miró, P.; Heine, T. *Adv. Mater.* **2016**, *28*, 853 – 856.

43  Bardeen, J.; Shockley, W. *Phys. Rev.* **1950**, *80,* 72 − 80.

44  Qiao, J.; Kong, X.; Hu, Z. X.; Yang, F.; Ji, W. *Nat. Commun.* **2014**, *5*, 4475.

45  Cai, Y.; Zhang, G.; Zhang, Y. W. *J. Am. Chem. Soc.* **2014**, *136,* 6269−6275.

46  Jeon, N. J.; Noh, J. H.; Kim, Y. C.; Yang, W. S.; Ryu, S.; Seok, S. *Nat. Mater.* **2014**, *13*, 897 − 903.

47  Shirayama, M.; Kadowaki, H.; Miyadera, T.; Sugita, T.; Tamakoshi, M.; Kato, M.; Fujiseki, T.; Murata, D.; Hara, S.; Murakami, T. N.; Fujimoto, S.; Chikamatsu, M.; Fujiwara, H. *Phys. Rev. Appl.* **2016**, *5*, 014012.

48  Blöchl, P. E. *Phys. Rev. B: Condens. Matter Mater. Phys.* **1994**, *50*, 17953 − 17979.

49  Kresse, G.; Joubert, D. *Phys. Rev. B: Condens. Matter Mater. Phys.* **1999**, *59*, 1758 − 1775.

50  Kresse, G.; Hafner, J. *Phys. Rev. B: Condens. Matter Mater. Phys.* **1993**, *47*, 558 − 561.

51  Perdew, J. P.; Burke, L.; Ernzerhof, M. *Phys. Rev. Lett.* **1996**, *77*, 3865 − 3868.

52  Grimme, S. *J. Comput. Chem.* **2006**, *27*, 1787 − 1799.